\documentclass{article}
\usepackage[T1]{fontenc} 
\usepackage[utf8]{inputenc} 
\usepackage{ismir,amsmath,cite,url}
\usepackage{graphicx}
\usepackage{color}
\usepackage{array}
\usepackage{caption}
\usepackage{subcaption}
\usepackage{xcolor}
\usepackage{makecell}
\usepackage{subcaption}
\usepackage{multirow,multicol,booktabs}
\usepackage{enumitem}

\newcommand{\AM}{\mbox{Stream}}
\newcommand{\AMone}{\mbox{Stream-1\%}}

\newcommand{\HR}{\mbox{HitRate}}
\newcommand{\NDCG}{\mbox{NDCG}}
\newcommand{\HardNeg}{\mbox{HardNeg}}
\newcommand{\GenreVRC}{VRC$_{\text{Genre}}$}
\newcommand{\ArtistVRC}{VRC$_{\text{Artist}}$}
\newcommand{\DeltaGenreVRC}{$\text{r\GenreVRC{}}^{(t)}$}

\newcommand{\CombinedGenre}[1]{$\lambda_{\text{Genre}}(#1)$}
\newcommand{\CombinedArtist}[1]{$\lambda_{\text{Artist}}(#1)$}
\newcommand{\LFM}{\mbox{LFM-1b}}

\usepackage{lineno}

\title{Multi-objective Hyper-parameter Optimization of Behavioral Song Embeddings}

\threeauthors
  {Massimo Quadrana} {Apple \\ {\tt mquadrana@apple.com}}
  {Antoine Larreche-Mouly} {Apple \\ {\tt alarreche@apple.com}}
  {Matthias Mauch} {Apple \\ {\tt mmauch@apple.com}}

\def\authorname{M. Quadrana, A. Larreche-Mouly, and M. Mauch}

\usepackage[bookmarks=false,pdfauthor={\authorname},pdfsubject={\papersubject},hidelinks]{hyperref}

\begin{document}

\maketitle
\begin{abstract}
Song embeddings are a key component of most music recommendation engines. 
%
In this work, we study the hyper-parameter optimization of behavioral song embeddings based on Word2Vec on a selection of downstream tasks, namely next-song recommendation, false neighbor rejection, and artist and genre clustering. We present new optimization objectives and metrics to monitor the effects of hyper-parameter optimization. We show that single-objective optimization can cause side effects on the non optimized metrics and propose a simple multi-objective optimization to mitigate these effects.
We find that next-song recommendation quality of Word2Vec is anti-correlated with song popularity, and we show how song embedding optimization can balance performance across different popularity levels.
We then show potential positive downstream effects on the task of play prediction.
Finally, we provide useful insights on the effects of training dataset scale by testing hyper-parameter optimization on an industry-scale dataset.


\end{abstract}
\section{Introduction}\label{sec:introduction}
Modern Recommendation Systems (RS) rely on embedding vectors to represent the latent user and item factors~\cite{pal20pinnersage}.
They can be employed for several downstream applications, ranging from song recommendation in radio stations or playlists~\cite{bonnin2014automated,chen18recsys_challenge,patwari2020semantically,irene19automatic,hansen21shifting,moore2012learning}, search~\cite{won2021multimodal,doh2021million}, tagging~\cite{ferraro2021enriched}, to the generation of artist and genre representations for annotation and recommendation tasks~\cite{chen2021learning,KorzeniowskiOG21,epure20multilingual}.
Embeddings are usually generated in the early stages of complex RS pipelines, often through self-supervised methods like Word2Vec~\cite{word2vec2013}, which come with default hyper-parameters tuned on non-RS tasks (e.g., NLP).

Practitioners and researchers in the field hence need feasible optimization objectives and reliable metrics to compare against when optimizing 
embeddings.
Recent research shows that hyper-parameter optimization can significantly improve recommendation quality~\cite{Caselles-Dupre2018,Twitter2020}. 
While prior work mainly focuses on recommendation tasks, it is worth considering other tasks like false positive rejection and clustering during optimization.

Additionally, embedding spaces are used as a source of knowledge and interpretation either through visualization tools~\cite{projector,tovstogan2020web}, or by using relationships between items from a statistical standpoint or by leveraging information outside the input data~\cite{predict,we_syntax,expressive,image_embe}.
We believe that music-RS practitioners and researchers could benefit from a deeper understanding of the behavior of song embedding optimization in relation to important factors such as song popularity. This could assist them in handling some emerging algorithmic biases like the popularity bias~\cite{abdollahpouri2019managing,kowald20_unfairness,vall2019order}.

\subsection{Contributions}
In this work, we provide a strong framework within which it is possible to monitor the performance of embedding models and evaluate the potential of the embedding model to adapt to new tasks. 
Our work presents a general methodology that can be applied to any embedding system and not only to Word2Vec.
The main contributions are:
\begin{itemize}[leftmargin=*]
  \item We define metrics and optimization objectives for three relevant tasks: next-song recommendation, false neighbor rejection, and genre and artist clustering.
  \item We demonstrate experimentally that single-objective optimization can have negative side effects on the non optimized metrics and propose a multi-objective optimization approach to combine recommendation and clustering objectives effectively.
  \item We show that next-song recommendation quality and song popularity are anti-correlated and reveal that song embedding optimization can balance performance across different popularity levels. 
  \item We show the potential positive downstream effects on the task of play prediction revealing that the benefits of song embedding optimization extend beyond the tasks considered during optimization. 
  \item Finally, we study embedding optimization at scale on an internal dataset of billions of listening events and show that increasing training dataset size allows for better configurations at the expense of longer optimization times.

\end{itemize}

\section{Method}

We consider song embeddings based on Word2Vec. 
This model was originally created to represent words in an English corpus through a self-supervised shallow neural network trained to learn dense vector representations of the words from sentences based on the surrounding words~\cite{word2vec2013}.
The same principle is now commonly used in recommender systems to compute item embeddings from user interaction sequences like site sessions or playlists~\cite{item2vec,in_your_inbox,quadrana17personalizing,vall2019order}. 

We study song embedding optimization with respect to four main tasks: next-song prediction, false neighbor rejection, artist clustering, and genre clustering. 
We define in this section the task and metrics that we used both for embedding optimization and evaluation of its effects on the resulting embedding space.

\subsection{Next-Song Prediction}
We choose next-song prediction as our primary target task. Next-item prediction is a common recommendation task whose goal is to predict the next item the user will interact with given some context~\cite{quadrana18_sequence}. In music recommenders, this often translates to predicting the next song given the history of songs played by the user~\cite{schedl19_dl_music}. 
To ensure the contextual relevance of recommendations, the past play history is often limited to the past few played songs or sessions, although more flexible solutions that look beyond a fixed horizon exist~\cite{hansen20_embeddings}.

Since our main goal is not to build the most accurate next-song predictor, but rather to measure effects of optimization on the embedding space as directly as possible, we simplified next-song prediction to the extreme case of predicting the next played song based only on the \emph{immediately preceding} song.
%
For each song in the evaluation set (the \emph{target} song henceforth), we consider the song the user played before it as the \emph{query} song.
For each (query, target) pair, we simply retrieve the top-100 exact nearest neighbors of the query and compute the average HitRate and Normalized Discounted Cumulative Gain (\NDCG{}) on the top-100 ranked neighbors. \HR{} is the fraction of times the target song is contained in the nearest neighbor of the query, while NDCG also accounts for the rank of the correct next-song within the predictions~\cite{voorhees2005trec}.

In order to ensure that improvements are due to true generalisation and not to overfitting, we split the evaluation into in-set and out-of-set.
In \emph{in-set evaluation}, we mask the last song in every training sequence and use second-to-last song as the query to compute next-song prediction metrics.
In \emph{out-of-set evaluation}, we hold-out the whole play sequences in the validation and test sets, and use every ordered pair of songs therein contained to compute the next-song prediction metrics.
We use the average in-set and out-of-set \HR{} and \NDCG{} values in our experiments.
The precise definition of a sequence varies between our experimental datasets and will be discussed in \secref{sec:datasets}.

\subsection{False Neighbor Rejection}
\begin{table}[t]
  \setlength{\tabcolsep}{2pt}
  \footnotesize
  \begin{tabular}{ll}
  \toprule
  Query song    & Hard Negative Neighbors                                                                     \\ \midrule
  \multirow{3}{*}{Paradise - Coldplay} & Snowman - Sia\\
  & Immigrant Song - Led Zeppelin\\
  & Basket Case - Green Day\\ 
  \midrule
  \multirow{3}{*}{Smells Like Teen Spirit - Nirvana} & Power - Kanye West\\ 
  & Ob-La-Di, Ob-La-Da - The Beatles\\ 
  & Rock Your Body - Justin Timberlake\\ 
  \midrule
  \multirow{3}{*}{Carry On Wayward Son - Kansas} & Natural - Imagine Dragons\\
  & Rap God - Eminen\\ 
  & It Ain't Me - Kygo \& Selena Gomez \\ 
  \bottomrule
  \end{tabular}
  \caption{Examples of hard negative neighbors in \AM{}.}
  \label{tab:hard_neg_examples}
  \vspace{-1em}
\end{table}

Since high quality nearest neighbors are essential to providing a good recommendations,
we are interested in evaluating the effectiveness of filtering out \emph{spurious} song neighbors, i.e., songs that are in the closest neighborhood of another song merely by chance and not due to real behavioral or metadata similarity.

To this end we define the task of False Neighbor Rejection.
We used a chi-squared $X^2$ test to identify false neighbors in the play sequence dataset~\cite{manning1999foundations}.
The $X^2$ test compares the observed co-occurrence frequencies of every pair of songs in the listening sequences against the expected co-occurrence frequencies in case of independence.
It is thus suitable to detect pairs of songs that appear in sequence by chance, and not because they are strongly related due to behavioral factors or metadata.
In practice, we compute the $X^2$ coefficient for each unordered bigram of songs in the play sequence dataset.
We consider as a \emph{hard negative neighbor} of a song any other song having high co-occurrence but chi-squared statistic below a significance threshold~\footnote{We empirically chose $10^{-7}$ times the sum of all song occurrences in the dataset as threshold.}.

One limitation of this approach is that it requires a large number of events to be able to find frequently co-occurring song pairs by chance. We were therefore able to run this analysis only on the \AM{} dataset, from which we retrieved 5M hard-negative pairs, with an average of $92$ hard-negatives per song.
Some examples of hard negative neighbors are shown in \tabref{tab:hard_neg_examples}.
We define the \HardNeg{} metric as the proportion of hard negatives for the top-100 nearest neighbors of every song (the smaller the better).
We will use it as a safeguard metric agains undesired side effects of song embedding optimization.

\subsection{Artist and Genre Clustering}\label{sec:tasks_clustering}


We are interested in measuring how strongly classes such as artists and genres cluster together in a given embedding space.
We introduce here the concept of Local Genre Coherence of the embedding space as the average fraction of songs in nearest neighbors set having the same primary genre as the query song. We similarly define the Local Artist Coherence as the average fraction of nearest neighbors belonging to the same artist as the query song.
For example, an embedding space has Local Genre Coherence $ = 0.5$ if on average $50\%$ of the nearest neighbors of each song have its same primary genre.

Computing Local Coherence metrics is a costly operation since it requires us to inspect the nearest neighbors of every embedded song. We instead propose to use as \emph{proxy metric} during optimization the much cheaper Cali{\'n}ski-Harabasz index or Variance Ratio Criterion (VRC)~\cite{variance_ratio_criterion} over artist and genre clusters in the embedding space.
The VRC is the ratio of the sum of between-clusters dispersion and of within-cluster dispersion for all genre clusters, where dispersion is the sum of squared distances.
Higher VRC values correspond to better separation among clusters, which is a desirable property because it reduces the chances of uncontrolled, cross-artist and cross-genre pollution in recommendations.

We study the optimization of embedding spaces with respect to clustering metrics and their relation to recommendation quality in \secref{sec:clustering} and \ref{sec:multiobj}.



\section{Datasets}
\label{sec:datasets}

\begin{table}[t]
  \footnotesize
  \centering
  \begin{tabular}{lccc}
    \toprule
  Dataset & \AM{} & \AMone{} & \LFM{} \\\midrule
  Songs                       & 66M                             & 17M & 31M                              \\
  Events                      & 255B                            & 2.6B & 1B                               \\
  Sequences                   & 0.9B                            & 9.3M & 120K                             \\
  Sequence length: mean & 277                        & 276 & 9044                 \\
  Sequence length: 25th percentile & 16                        & 16 & 1138                 \\
  Sequence length: median & 29                        & 27 & 3930                 \\
  Sequence length: 75th percentile & 58                        & 57 & 16155                 \\

  \bottomrule
  \end{tabular}
  \caption{The statistics of the datasets.}
  \label{tab:datasets}
  \vspace{-1em}

  \end{table}

We run our experiments on two datasets. 
The first is a large-scale proprietary dataset of anonymized streaming listening sequences and playlists. 
We call this dataset \AM{}. 
The second is the \LFM{} dataset which contains 1B time-stamped listening events collected from Last.fm~\cite{scheld16_lfm_dataset}.

The main statistics of each dataset are shown in \tabref{tab:datasets}. The sequence length distribution differs significantly between the two datasets. \AM{} contains large numbers of both long listening suquences and short playlists (the median length is 58, but mean is 277), while \LFM{} contains mainly long listening sequences (the median length is 3930).
This will have an impact on the optimal hyper-parameters discovered for each dataset.

In order to compute artist and genre clustering metrics we need song-level artist and genre annotations.
For the \AM{} dataset, each song in is mapped to the corresponding primary genre and artist through an internal auto-tagging pipeline.
\LFM{} already comes with song to artist mappings, however song-level genre annotations are not directly available. We hence mapped each song to the \emph{first} Freebase artist genre from the LFM-1b User Genre Profile dataset~\cite{schedl17_lfm_ugp_dataset}. We are aware of the noise introduced by this approximation, yet we believe it provides a useful contribution towards the reproducibility of our experiments.

We partition both datasets by randomly sampling sequences without replacement with proportions 98/1/1 for training, validation and test respectively.
However, \AM{} still contains $7500$ times more sequences than \LFM{}, which makes optimization at full-scale impractical. We hence downsample the sequences in the dataset using a 1\% rate. We will refer to this subsampled dataset to as \AMone{}. Scaling up optimization to the full dataset is a challenging task that we will investigate in more detail in \secref{sec:multiscale}.

\section{Experiments}
\begin{table}[t]
  \footnotesize
  \centering
  \begin{tabular}{cccccc}
  \toprule
  Parameter & Default & Range              & Description                \\
  \midrule
  $d$       & $100$     & $[25, 200]$     & Embedding vector dimension \\
  $L$       & $5$       & $[1, 40]$     & Sliding window max length  \\
  $\alpha$  & $0.75$    & $[-1.0, 1.0]$   & Negative sampling exponent \\
  $N$       & $5$       & $[1, 100]$     & Number of negative samples \\
  $\lambda$ & $0.025$   & $[0.001, 0.1]$  & Initial learning rate \\
  \bottomrule     
  \end{tabular}
  \caption{Default Word2Vec hyper-parameters and their respective optimization ranges.}
  \label{tab:params}
  \vspace{-0.5em}
\end{table}

\begin{table*}[t]
  \centering
  \footnotesize
  \begin{tabular}{@{}llccccccccc@{}}
    \toprule
  & & \multicolumn{5}{c}{\textbf{\AMone{}}} & \multicolumn{4}{c}{\textbf{\LFM{}}}\\
  \addlinespace[3pt]
  Opt. Type & Objective & \HR{} & \NDCG{} & \HardNeg{} & \GenreVRC{} & \ArtistVRC{} & \HR{} & \NDCG{} & \GenreVRC{} & \ArtistVRC{}  \\
  \cmidrule(r){1-2} \cmidrule(lr){3-7} \cmidrule(l){8-11} \\
  N/A &   N/A & 0.3538 & 0.1378 & 0.0180 & 927 & 3.373 & 0.3771 & 0.1562 & 520 & 4.141  \\ 
  \cmidrule(r){1-2} \cmidrule(lr){3-7} \cmidrule(l){8-11} \\
  \multirow{3}{*}{Single-obj} & \HR{} & 0.3725$^\uparrow$ & 0.1482$^\uparrow$ & 0.0108$^\uparrow$ & 1033 & 4.437 & 0.4492$^\uparrow$ & 0.2050$^\uparrow$ & 382 & 3.457  \\
  & \GenreVRC{} & 0.3000 & 0.1218 & 0.0146 & 2006 & 5.521 & 0.3776 & 0.1609 & 1293 & 8.603  \\
  & \ArtistVRC{} & 0.3402 & 0.1336 & 0.0175 & 1397 & \textbf{34.347} & 0.3532 & 0.1488 & \textbf{1444} & \textbf{94.476}  \\ 
  \cmidrule(r){1-2} \cmidrule(lr){3-7} \cmidrule(l){8-11} \\
  \multirow{4}{*}{Multi-obj} & \CombinedGenre{0.01} & \textbf{0.3772}$^\uparrow$ & \textbf{0.1586}$^\Uparrow$ & 0.0084$^\uparrow$ & 1495 & 5.216 & 0.4428$^\uparrow$ & \textbf{0.2092}$^\uparrow$ & 623 & 5.955  \\
  & \CombinedGenre{0.1} & 0.3742$^\uparrow$ & 0.1515$^\Uparrow$ & 0.0096$^\uparrow$ & 1617 & 5.771 & 0.4067$^\uparrow$ & 0.1698$^\uparrow$ & 997 & 7.298  \\
  & \CombinedArtist{0.01} & 0.3699$^\uparrow$ & 0.1520$^\Uparrow$ & \textbf{0.0057}$^\Uparrow$ & 1166 & 4.580 & 0.4458$^\uparrow$ & 0.1883$^\uparrow$ & 487 & 5.469  \\
  & \CombinedArtist{0.1} & 0.3331 & 0.1332 & 0.0093 $^\uparrow$ & \textbf{2233} & 7.422 & \textbf{0.4537}$^\uparrow$ & 0.1998$^\uparrow$ & 619 & 6.022  \\ 
  \bottomrule
  \end{tabular}
  \caption{Results of hyper-parameter optimization on both datasets. The first line reports the metrics for the default configuration. Best results are in \textbf{bold}. For HR/NDCG/HardNeg only: $^\uparrow$ and $^\Uparrow$ denote stat.~sig.~improvement over the default and the best single-objective configurations respectively (paired t-test at $p < 0.01$ with Bonferroni correction).}
  \label{tab:hpo-res}
  \vspace{-0.5 em}
  \end{table*}

We report here the optimization of song embeddings for the tasks defined in the previous section. 
We optimize the hyper-parameters of skip-gram Word2Vec by running Bayesian Hyper-Parameter Optimization (HPO)~\cite{snoek2012practical}, initialized with 10 iterations of Random Search~\cite{bergstra2011algorithms} before running Bayesian Search until convergence. Similarly to previous work, we constrained all training times to be approximately equal to the default Word2Vec configuration~\cite{Twitter2020}. This ensures a fair comparison among trials, and prevents the optimizer from discovering configurations that are impractical to train at large scales.

\subsection{Background on Word2Vec}
Both variants of Word2Vec, Skipgram and the Continuous Bag of Words (CBOW), are self-supervised shallow neural network models trained by minimizing the categorical cross-entropy loss with approximation softmax~\cite{word2vec2013}.
Here we consider Skipgram with negative sampling for its superior computational efficiency~\cite{mnih2012fast}.

The hyper-parameters of Word2Vec, their defaults and optimization ranges are detailed in \tabref{tab:params}. In short, $d$ is the embedding size, $L$ is the maximum window length, $\alpha$ controls the negative sampling (uniform sampling for $\alpha=0$, popularity sampling $\alpha>0$, inverse popularity sampling for $\alpha<0$), $N$ is the number of negative samples used to approximate the softmax and $\lambda$ is the learning rate.

\subsection{Optimizing for Next-Song Recommendation}
We first analyze the optimization of next-song recommendation quality by running HPO with the \HR{} objective. In line with previous works, we observe significant improvements with respect to the default configurations on both the tested datasets (second line of \tabref{tab:hpo-res}).
On \AMone{}, \HR{} improves by 5\% and NDCG by 7\%, while \HardNeg{} reduces drastically by 40\%. The reduction in \HardNeg{} ensures that the improvement in recommendation accuracy does not comes at the expense of more false positive neighbors. 
On \LFM{} we similarly observe +19\% \HR{} and +31\% \NDCG{}.
The results on this task are in line with previous findings~\cite{Twitter2020,Caselles-Dupre2018}.

We are now also able to monitor the effects of optimization on Genre and Artist Clustering metrics. While, \GenreVRC{} and \ArtistVRC{} slightly increase on \AM{}, they both reduce on \LFM{}. This suggests that recommendation and clustering metrics may be slightly anti-correlated for this dataset. We will investigate this effect further in \secref{sec:clustering} and \ref{sec:multiobj}.

\subsection{Optimising for Genre and Artist Clustering}\label{sec:clustering}
\begin{figure}[b]
  \vspace{-1em}
  \centering
  \begin{subfigure}[t]{0.49\columnwidth}
      \includegraphics[width=\columnwidth]{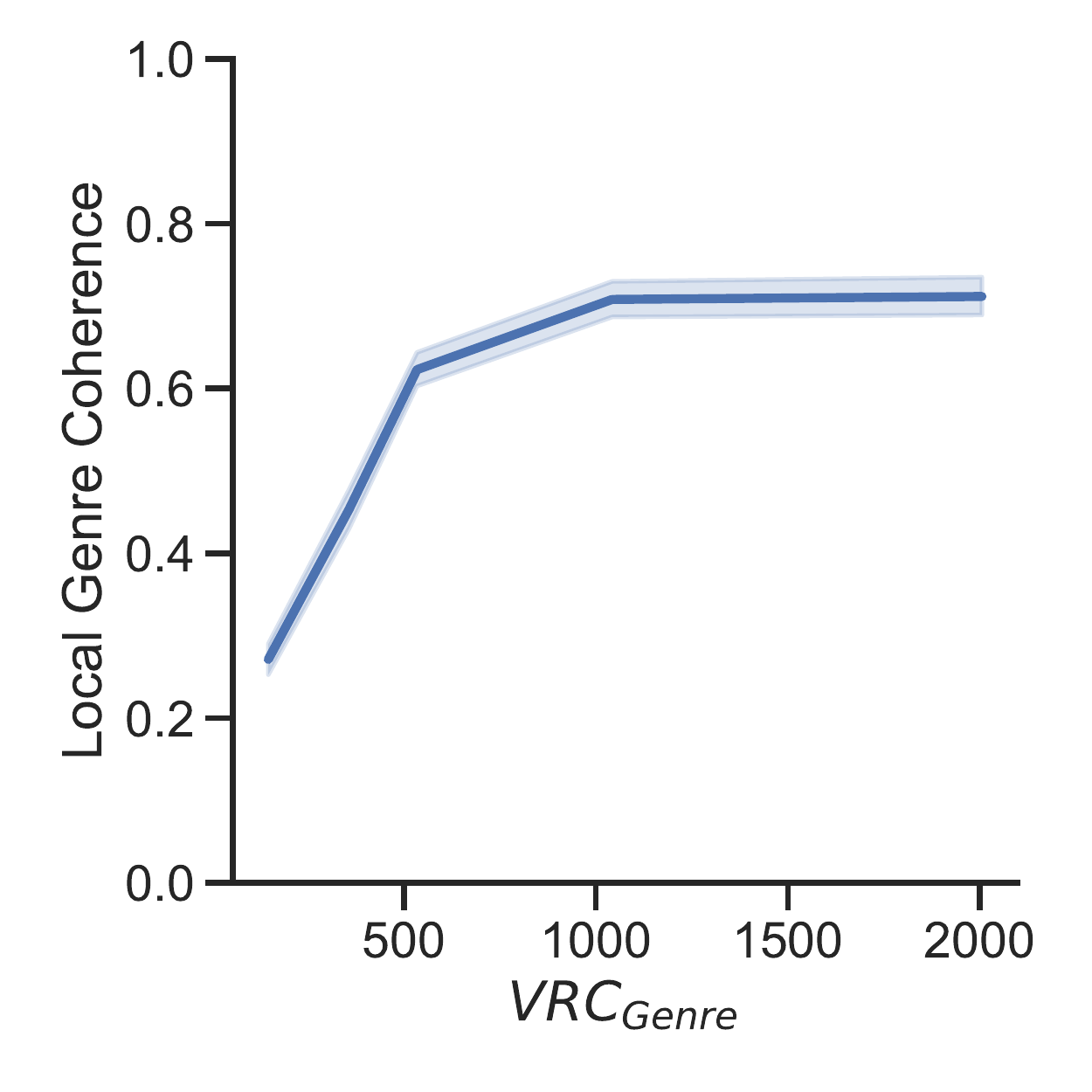}
  \end{subfigure}
  \hfill
  \begin{subfigure}[t]{0.49\columnwidth}
     \includegraphics[width=\columnwidth]{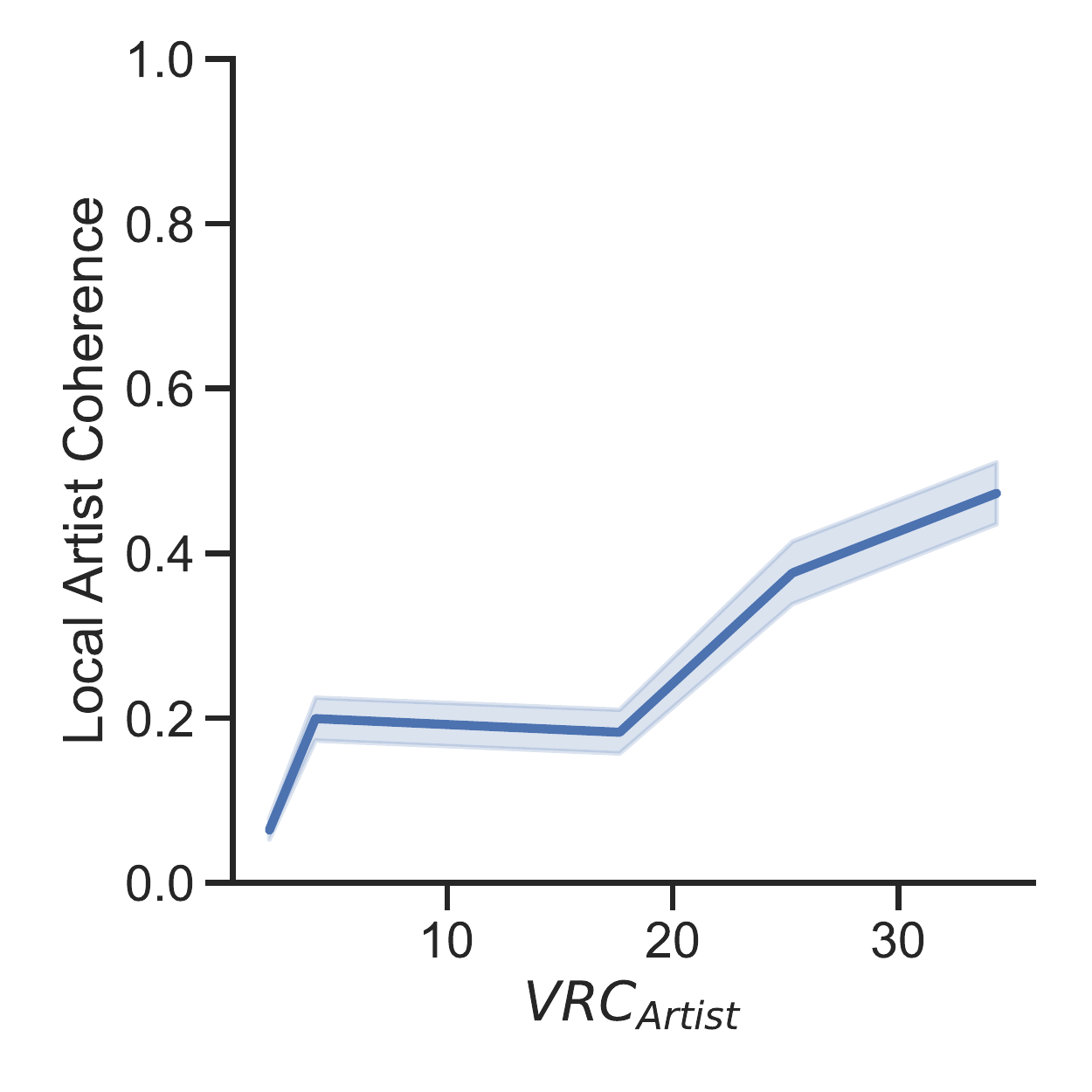}
  \end{subfigure}
  \caption{Local Coherence for genres (left) and artists (right) on \AMone{}. Means and 95\% Confidence Intervals are shown.}
  \label{fig:vrc_to_local_coherence}
  \vspace{-1em}

\end{figure}

We now analyze the optimization of embeddings with respect to the Local Genre and Artist Coherence by using the  Variance Ratio Criterion as a proxy objective. 

We first show that VRC is a suitable proxy by comparing it against Local Coherence metrics on \AMone{}.
We computed Local Genre Coherence on 500 songs with at least 10k plays selected using stratified sampling across the top-10 played genres and using 50 nearest neighbors per song. \figref{fig:vrc_to_local_coherence} (left) shows strong positive correlation between average Local Genre Coherence and \GenreVRC{} of 5 embedding spaces generated with different hyper-parameters. 
%
%
We similarly computed Local Artist Coherence by sampling 125 artists having at least 25 songs using stratified sampling by artist popularity to account for popularity biases. We then average artist coherence score over 5 songs per artist and 50 nearest neighbors per song.
\figref{fig:vrc_to_local_coherence} (right) shows positive correlation between Local Artist Coherence and \ArtistVRC{} of 5 embedding spaces generated with different hyper-parameters. 

\tabref{tab:hpo-res} shows that the optimal configuration for genre clustering on \AMone{} doubles the \GenreVRC{} score compared to the default hyper-parameters, but with a significant reduction in terms of \HR{} and \NDCG{}. The next-song recommendation quality is thus badly affected by the myopic optimization of genre coherence.
This effect is slightly less evident for artist clustering optimization, in which we were able to obtain 10 fold greater \ArtistVRC{} with negligible decrease in \HR{} and \NDCG{}. In both cases, HardNeg is not significantly affected.

We observe similar effects on \LFM{}, where significant improvements in \GenreVRC{} and \ArtistVRC{} correspond to non significant improvement (genre clustering) or significant deterioration (artist clustering) of next-song recommendation metrics.

For both datasets, the optimal configurations have significantly worse next-song recommendation quality than the optimal one found by single-objective next-song recommendation. This suggests that optimizing embeddings for clustering alone can seriously harm the recommendation quality. In the next section we tackle the problem of optimizing both objectives simultaneously.

\subsection{Multi-objective Optimization}\label{sec:multiobj}
High quality next-song recommendation and genre/artist clustering are both desirable properties of the embedding space but, as we have just observed, optimizing for a single objective can harm others.
We investigate here the simultaneous optimization of both objectives through Multi-Objective Hyper-parameter Optimization (MOHPO).

The simplest approach to MOHPO is scalarization, which transforms a multi-objective goal into a single-objective one. Examples of scalarization are the weighted sum, Tchebycheff or $\epsilon$-constraint approaches~\cite{karl22multi}.
Advanced MOHPO techniques extend evolutionary algorithms and Bayesian Optimization to explicitly handle different trade-offs between the multiple objectives~\cite{karl22multi}. The choice of the optimal MOHPO method is beyond the scope of this work, hence we opted for Bayesian Optimization with scalarization since it fits easily into our existing optimization framework.

We focus on the joint optimization of \HR{} and \GenreVRC{}. The same reasoning holds for \HR{} and \ArtistVRC{} and it is omitted for space reasons. Since these metrics have widely different scales, we first define the relative improvement in \GenreVRC{} of a new hyper-parameter configuration $t$ with respect to the default configuration:
\begin{equation}
  \text{\DeltaGenreVRC{}} = \frac{\text{\GenreVRC{}}^{(t)} - \text{\GenreVRC{}}^{(def)}}{\text{\GenreVRC{}}^{(def)}}
\end{equation}
where $\text{\GenreVRC{}}^{(def)}$ is the \GenreVRC{} for the default configuration.
The new scalarized objective is the simple convex combination of the two objectives:
\begin{equation}
  \text{\CombinedGenre{\alpha}} = \alpha \text{\HR{}}^{(t)} + (1 - \alpha) \text{\DeltaGenreVRC{}}
\end{equation}
where $\alpha$ controls the relative weight of next-song recommendation and genre clustering objectives in the optimization. 

We tried values of $\alpha = 0.01$ and  $0.1$.
Results shown in \tabref{tab:hpo-res} highlight the effectiveness of this approach on \AMone{}. 
While there is not a single solution that performs best on all metrics, the configuration found with objective \CombinedGenre{0.01} \emph{dominates} the best solution discovered by next-song optimization on all metrics, with statistically better NDCG (+7\%). The configuration found with objective \CombinedArtist{0.01} also achieves statistically better HardNeg, with a reduction of 68\%, at the expense of lower clustering scores.
In both cases increasing the value of $\alpha$ to $0.1$ we can effectively trade some next-song prediction accuracy for better genre and artist clustering quality.
The optimal setup depends on the final application.

Similarly, on \LFM{} we observe that the best next-song recommendation strategies are found through multi-objective optimization, although no solution entirely dominates any single-objective this time. This fact can be partly attributed to the noisy genre annotations we have available for this dataset.

These results highlight the effectiveness of combining recommendation and clustering objectives, which can mutually benefit from each other if combined properly.

\subsection{Insights on Song Popularity}

The effects of popularity on recommendations are subject of intense research activity within the research community~\cite{abdollahpouri2019managing,kowald20_unfairness,vall2019order}.
We contribute here by studying the effects of next-song recommendation HPO on (query, target) song pairs belonging to various buckets of popularity.

We first categorize songs into buckets of popularity, each of which comprises 20\% of the total listening events in the dataset, being $0$ the smallest bucket with most popular songs and $4$ the largest bucket containing the least popular ones\footnote{The songs that account for the top-20\% of all plays end in bucket $0$, the ones for the top-20\% to 40\% end in bucket $1$, etc.}.
We randomly sample 1k songs per bucket pair and the aggregate \HR{} based on query \emph{and} target bucket. For the sake of space, we analyze the \AM{} dataset only here. The results on \LFM{} are available in the Supplementary Material.

\figref{fig:am_heat_hr_prod} shows the values of \HR{} by bucket pair for the default configuration of Word2Vec.
We first observe that the recommendation quality is localized to the same or nearest popularity bucket to which the query song belongs.
As the query and target begin to differ in popularity, \HR{} drops.
We clearly see that \HR{} is anti-correlated with popularity.
One possible explanation is that popular songs by definition give less information about the user tastes and hence have larger sets of plausible ``next-songs''. On the other hand, the less popular songs often belong to taste ``niches'' and are thus easier to model.

\figref{fig:am_heat_hr_tuned} shows the values of \HR{} by bucket pair for embeddings for the optimal \CombinedGenre{0.01} multi-objective configuration.
The behavior of localized performance and anti-correlation observed using the default configuration is still visible.
However, we notice a significant difference between the two configurations across the main diagonal.
We have highlighted this difference in \figref{fig:am_hr_quantiles_bars}.
We can see that tuning brings significant gains in \HR{} at all buckets except for the least popular songs in $4$. We do not observe the same performance drop on \LFM{} though (see the Supplementary Material). 
On both datasets optimization seems to balance recommendation accuracy more across popularity buckets, which is a desirable effect.

\begin{figure}[t]
  \centering
  \begin{subfigure}[t]{0.31\columnwidth}
      \includegraphics[width=\textwidth]{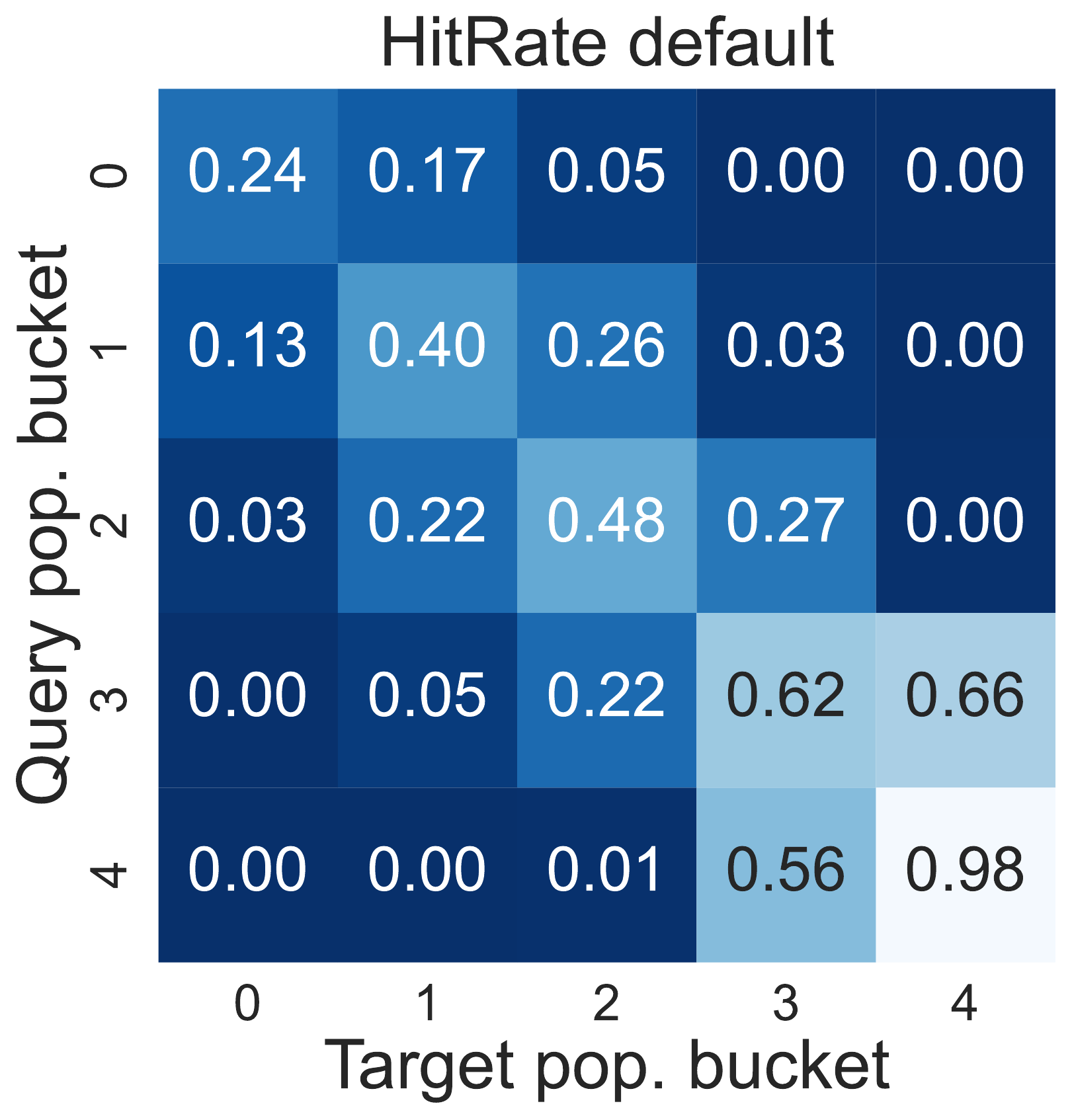}
      \caption{}
      \label{fig:am_heat_hr_prod}
    \end{subfigure}
  \hfill
  \begin{subfigure}[t]{0.31\columnwidth}
      \includegraphics[width=\textwidth]{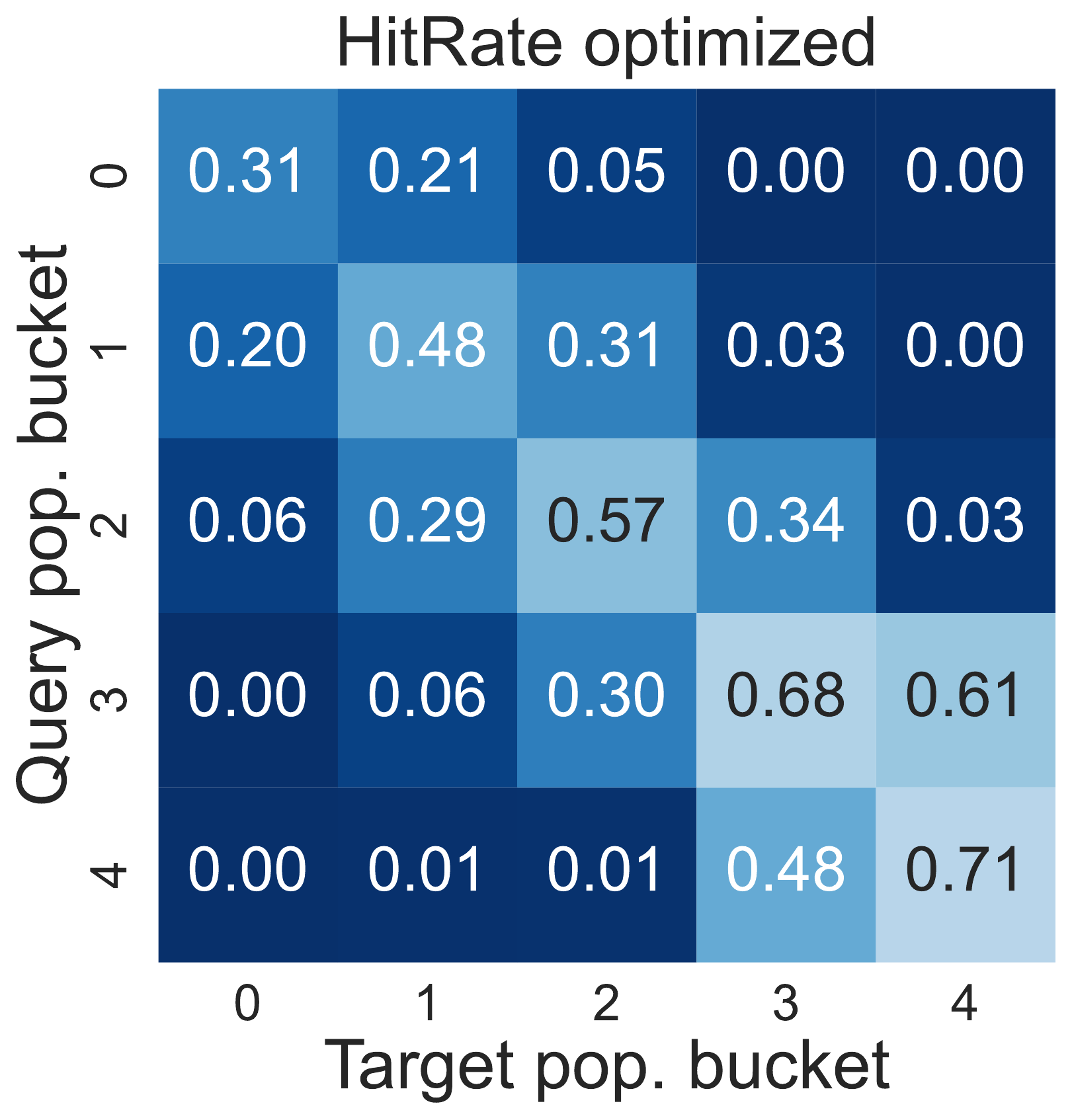}
      \caption{}
      \label{fig:am_heat_hr_tuned}
    \end{subfigure}
  \hfill
  \begin{subfigure}[t]{0.31\columnwidth}
      \includegraphics[width=\textwidth]{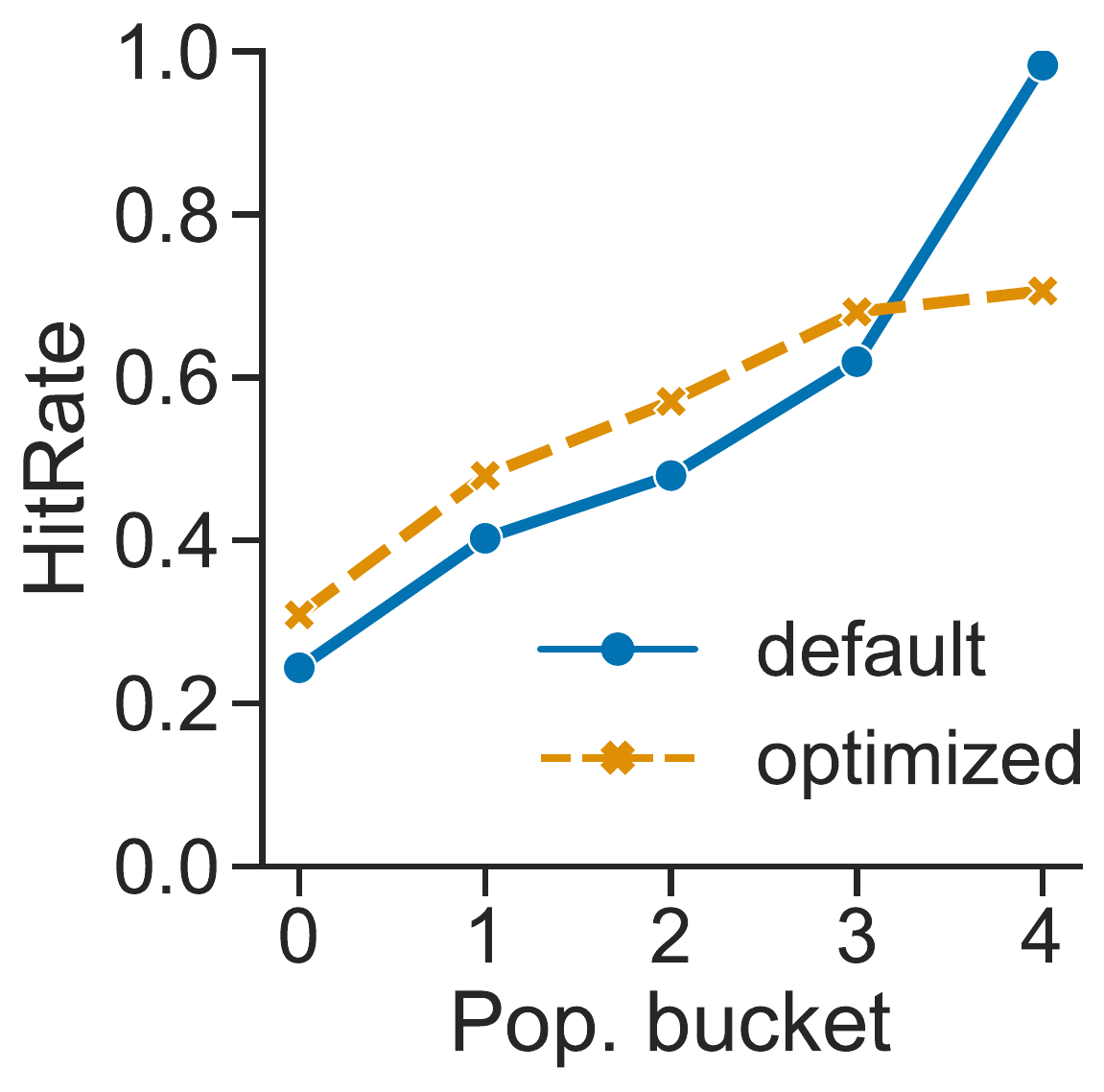}
      \caption{}
      \label{fig:am_hr_quantiles_bars}
    \end{subfigure}
  \caption{Query-target popularity analysis for \AM{}:~\HR{}~by popularity bucket for the default configuration~(a)~and the optimized one~(b), and \HR{}~for song pairs in the same popularity bucket (c). Bucket $0$ contains the most popular songs.}
  \vspace{-0.5em}
 \end{figure}



\subsection{Insights on Play Prediction}

As song embeddings are usually used as inputs to other pipelines, we wanted to explore whether embedding optimization had beneficial effects on a different task from the one it was originally run on.
For the sake of this experiment, we considered the problem of Play Prediction in seeded radio and autoplay.
In seeded radio, the user selects a \emph{seed}, e.g., an artist or a song, and the recommender generates an endless sequence of songs that are related to that seed.
Similarly, autoplay generates a stream of music starting from the last played song in an album or a playlist.
Seeded radio and autoplay are two prominent product features that allow users to generate streams of songs that are fully machine-learning driven.
In both cases, the user organically selects only the seed or the collection from which to start the stream of music, and everything else is left to the recommendation algorithm to decide.

In this context, we study the reaction that users have on the \emph{first} recommended song, since it is where the transition from organic to algorithmic selection happens.
A bad listening experience at this point could easily induce users to stop listening entirely.
We thus consider the task of predicting whether the first song generates a play event that lasts at least 30 seconds.
We call this task Play Prediction.

Building an accurate predictor is beyond the scope of this paper. We instead measure the correlation between the Play Rate, i.e., the average number of times one organic play is followed by a successful algorithmic play, and the cosine similarity between the embeddings of the recommended and the seed songs.
We use Word2Vec embeddings trained on the full \AM{} dataset. We compare default hyper-parameter configuration to the single-objective and multi-objective configurations having the highest \HR{} on this dataset. We compute Play Rates on a proprietary dataset composed of 4.2M song pairs extracted from 184M listening sessions.

\begin{figure}[t]
  \centering
  \includegraphics[width=0.52\columnwidth]{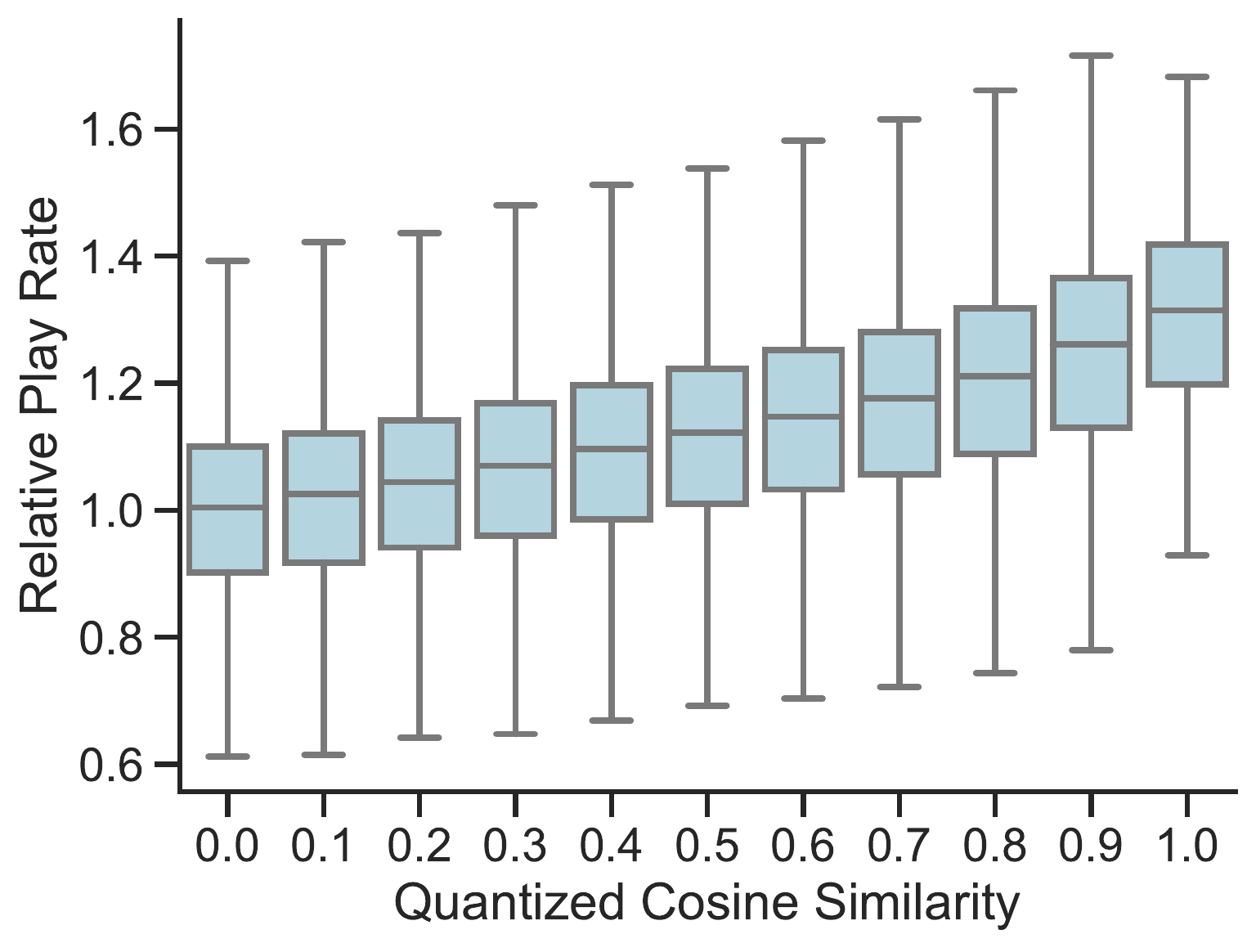}
  \caption{Visualization of the positive correlation between Relative Play Rate and  cosine similarity on default embeddings (similarities quantized to the first decimal). Play Rates are scaled relative to cosine similarity $ = 0.0$ to preserve business sensitive values.}
  \label{fig:royalty_cosine}
  \vspace{-0.5em}
\end{figure}

\figref{fig:royalty_cosine} shows that there is a strong positive correlation between Play Rate and cosine similarity between embeddings.
We observe positive correlation for both the default and optimized embedding spaces (\tabref{tab:correlation}).
The correlation is stronger for the most frequent pairs ($\ge 100$ occurrences).
However, the correlation is stronger for the optimized embeddings than the default ones, $+4\%$ for the best single-objective configuration and $+12\%$ for the best multi-objective configuration. 
This suggests that embedding optimization can have beneficial effects on tasks different from the one it was initially designed to address. Furthermore, it provides additional evidence on the superiority multi-objective optimization over single-objective one.
 
\begin{table}[t]
  \footnotesize
  \centering
  \begin{tabular}{lccc}
  \toprule
    & Default & Single-obj & Multi-obj\\
  \midrule
  All pairs & 0.2055  & 0.2140 & 0.2304 \\
  Frequent pairs & 0.3335  & 0.3480 & 0.3717 \\
  \bottomrule
  \end{tabular}
  \caption{Pearson's correlation between Play Rate and cosine similarity for default and optimized embeddings.}
  \vspace{-1.5em}
  \label{tab:correlation}
  \end{table}

  \subsection{Optimization at Scale}\label{sec:multiscale}
  
  \begin{table}[t]
    \setlength{\tabcolsep}{3pt}
    \footnotesize
    \centering
    \begin{tabular}{@{}ccccccc@{}}
      \toprule
       Sampling& Opt.~time& \HR{} & \NDCG{} & \HardNeg{} & \GenreVRC{} & \ArtistVRC{} \\
       \midrule
    N/A & N/A & 0.3079 & 0.1217 & 0.0126 & 7229 & 3.266 \\
    1\% & 28h & 0.3432 & 0.1378 & 0.0109 & 7367 & 3.780 \\
    2\% & 47h & 0.3499 & 0.1421 & 0.0063 & 6886 & 3.374 \\
    5\% & 83h & 0.3729 & 0.1506 & 0.0167 & 10410 & 3.992 \\
    \bottomrule
    \end{tabular}
    \caption{Results of multiple scale HPO on \AM{}, with the total optimization time and the best metrics obtained full scale (first line refers to the default configuration). Better performance is obtained at the largest subsample rates at the expense of longer optimization times. All pairwise comparisons on \HR{}, \NDCG{} and \HardNeg{} are stat.~sig. at $p < 0.01$ using Bonferroni correction, except for \HardNeg{} between default and 1\% sampling. }
    \label{tab:scale}
    \vspace{-1em}
    \end{table}

  To the best of our knowledge, there exists little literature on hyper-parameter optimization over massive datasets with billions of sequences and events.
  Previous work on the topic either optimize directly on the full dataset, which is unfeasible at scales like our \AM{} dataset, or consider a \emph{fixed} subsample rate~\cite{Twitter2020}.
  However, it is unclear which subsample rate would lead to results that are representative of performance at full-scale.
  
  We explore this aspect by running Hyper-Parameter Optimization at increasing training dataset scales. The best hyper-parameters found at each data scale are then used to train the model at full-scale.
  We limit training time to stay within 25\% more the time of training runs with the default hyper-parameters.
  This allows us to identify the best representative subsample rate while keeping optimization times within reasonable limits.
  
  We split the \AM{} dataset into 4 overlapping training sets containing 1\%, 2\%, 5\% and 98\% of randomly selected sequences respectively, We run Bayesian hyper-parameter search on the first 3 splits and use the best hyper-parameters from each run to train Word2Vec at full 98\% scale. For simplicity, we consider just single-objective \HR{} optimization.
  \tabref{tab:scale} shows that there is significant correspondence between the sampling rates used during optimization and the final performance on the full scale dataset. The optimal configuration found at 5\% scale is by far the best when evaluated at full-scale on all metrics. However, the total optimization time increases significantly when larger subsamples are used.

\section{Conclusions}
In this paper we analyzed the offline optimization of song embeddings through the lens of the tasks they are often employed on downstream. We proposed an effective way to optimize Word2Vec hyper-parameters on recommendation and clustering tasks jointly, with substantial benefits over their respective single-objective optimization variants. We investigated the interactions between next-song recommendation and song popularity. Our results suggest that careful optimization has the desirable property of balancing algorithm performance across popularity buckets. We showed the potential positive effects of optimization on the downstream task of Play Prediction, which provided further evidence on the superiority of multi-objective optimization over single-objective one. Finally, we investigated the effects of optimization at scale, which is particularly relevant for industrial applications.

As future work, we plan to validate these insights with online A/B testing. Our approach can be extended to other tasks, like diversity and fairness, and to the latest findings in Multi-Objective Hyper-parameter Optimization~\cite{majid19multiobjective,karl22multi}. 

\section{Aknowledgements}
We would like to thank Matt Jockers and all the members of the Music Machine Learning team for their help in improving and proofreading this work.

\bibliography{ISMIRtemplate}

\section{Supplementary Material}

\subsection{Query-popularity analysis on \LFM{}}
\begin{figure}[h]
  \centering
  \begin{subfigure}[t]{0.31\columnwidth}
    \includegraphics[width=\textwidth]{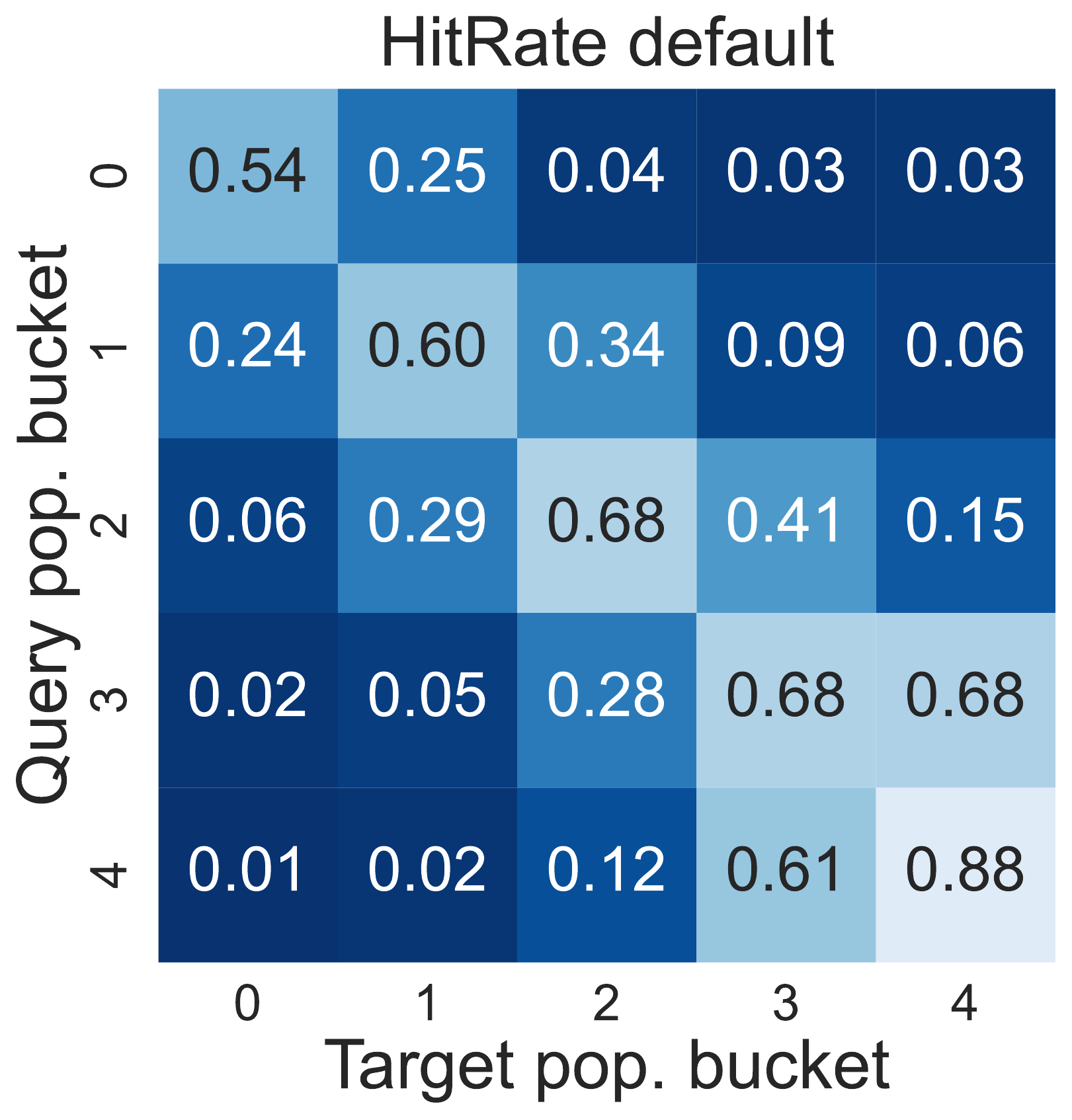}
    \caption{}
    \label{fig:lfm_heat_hr_prod}
  \end{subfigure}
  \hfill
  \begin{subfigure}[t]{0.31\columnwidth}
    \includegraphics[width=
\textwidth]{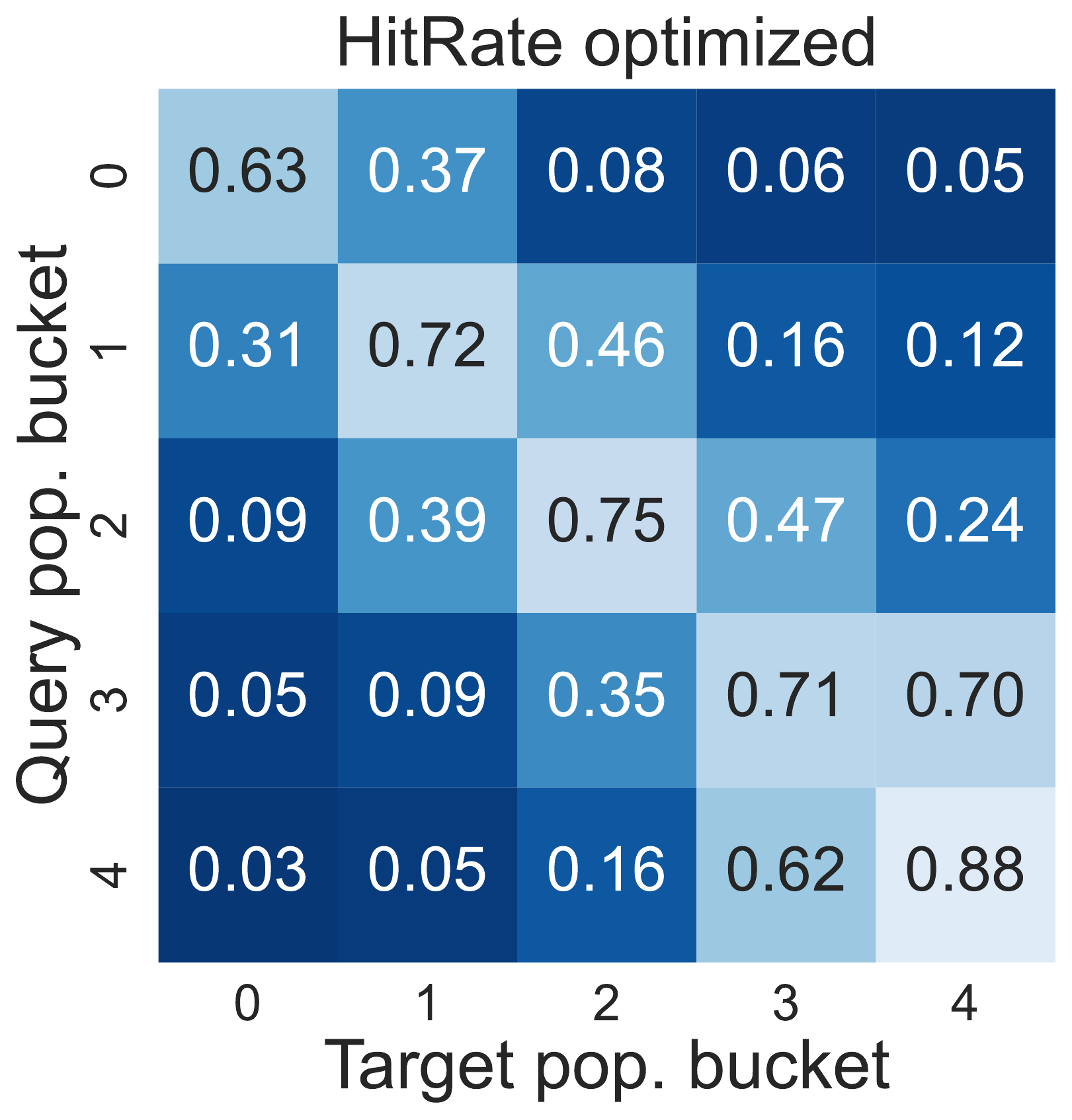}
    \caption{}
    \label{fig:lfm_heat_hr_tuned}
  \end{subfigure}
  \hfill
  \begin{subfigure}[t]{0.31\columnwidth}
    \includegraphics[width=\textwidth]{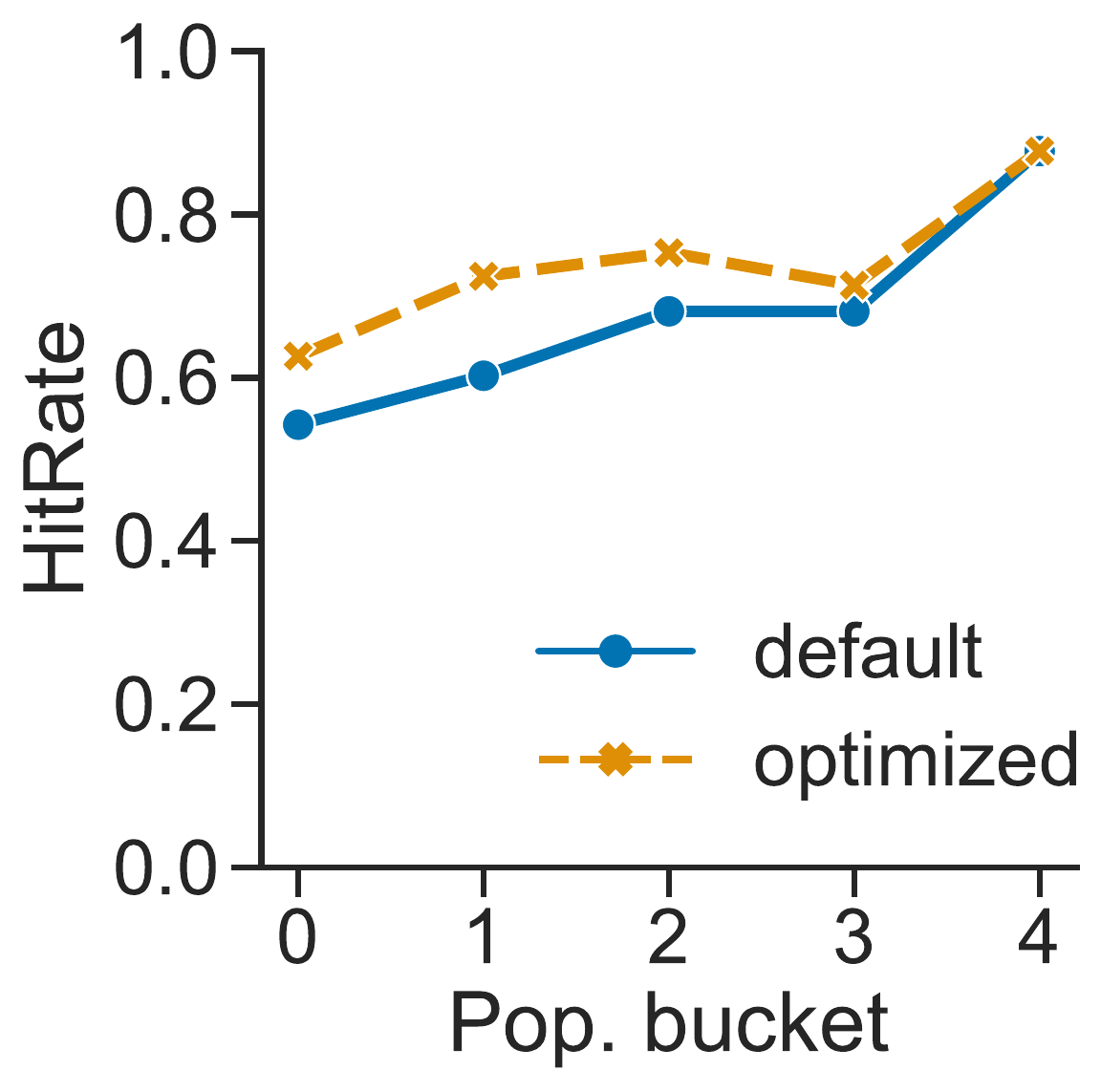}
    \caption{}
    \label{fig:lfm_hr_quantiles_bars}
  \end{subfigure}
  \hfill
  \caption{Query-target popularity analysis for \LFM{}: HR by quintile for the default configuration (a) and optimized one (b), and HR for song pairs in the same quantile (c).}
 \end{figure}
 
\end{document}